\documentclass[12pt]{iopart}

\usepackage{cite}
\usepackage{dcolumn}
\usepackage{bm}
 \expandafter\let\csname equation*\endcsname\relax
  \expandafter\let\csname endequation*\endcsname\relax
\usepackage{amsmath}
\usepackage{braket}
\usepackage{color}
\usepackage{float}
\usepackage{graphicx}
\usepackage[caption=false]{subfig}

\newcommand{\ud}{\mathrm{d}}
\renewcommand{\mr}{\mathbf{r}}

\newcommand{\mG}{\mathbf{G}}

\newcommand{\mE}{\mathbf{E}}
\newcommand{\mH}{\mathbf{H}}

\newcommand{\mF}{\tilde{\mathbf{F}}}
\newcommand{\mft}{\tilde{\mathbf{f}}}
\newcommand{\tlo}{\tilde{\omega}}

\newcommand{\mn}{\bf n}

\definecolor{myBlue}{rgb}{0.0430,0.5156,0.7773}
\definecolor{light-gray}{gray}{0.95}

\newcommand{\rev}[1]{{\color{blue}#1}}

\begin{document}

\abovedisplayskip=8pt
\abovedisplayshortskip=0pt
\belowdisplayskip=8pt
\belowdisplayshortskip=8pt


\title[QNM approach to modelling light-emission and propagation in nanoplasmonics]{Quasinormal mode 
approach to modelling light-emission and propagation in nanoplasmonics}

\author{Rong-Chun Ge$^1$, Philip Tr{\o}st Kristensen$^2$, Jeff. F. Young$^3$,   and Stephen Hughes$^{1}$}
\address{
$^1$Department of Physics, Engineering Physics and Astronomy, Queens University, Kingston, Ontario, Canada K7L 3N6\\
$^2$DTU Fotonik, Technical University of Denmark, DK-2800 Kgs. Lyngby, Denmark\\
$^3$Department of Physics and Astronomy, University of British Columbia, 6224 Agricultural Rd., Vancouver,
British Columbia, Canada V6T 1Z1}
\ead{shughes@physics.queensu.ca}

\begin{abstract}
{We describe a powerful and intuitive technique for modeling light-matter interactions in classical and quantum nanoplasmonics. Our approach uses a  quasinormal mode expansion of the Green function within a metal nanoresonator of arbitrary shape, together with a Dyson equation, to derive an expression for the spontaneous decay rate and far field propagator from dipole oscillators outside resonators. For a single quasinormal mode, at field positions outside the quasi-static coupling regime, we give a closed form solution for the Purcell factor and generalized effective mode volume. We augment this with an analytic expression for the divergent LDOS very near the metal surface, which allows us to derive a simple and highly accurate expression for the electric field outside the metal resonator at distances from a few nanometers  to infinity. 
This intuitive formalism  provides an enormous simplification over full numerical calculations and fixes several pending problems in quasinormal mode theory.
}

\end{abstract}

\pacs{42.50.Pq, 78.67.Bf, 73.20.Mf}
\maketitle

\section{Introduction}
High-index-contrast dielectric cavities and metallic nano-resonators  both facilitate control of light-matter interaction by engineering the local density of {optical} states (LDOS). While this occurs on the scale of a wavelength for dielectric cavities \cite{cavityBook}, it can extend down to nm length scales for metallic nanoparticles (MNPs) \cite{plasmonBook}. Traditional quantum optical effects such as vacuum Rabi oscillations and resonance fluorescence   become much  richer with MNPs~\cite{Hohenester:PRB08,Vanvlack:PRB12,Hughes13}. Localized surface plasmons dramatically increase the LDOS within a few nm of a MNP, which can enhance the spontaneous emission (SE) rate of a single photon emitter~\cite{Yao10,Novotny06,Frimmer13}. The pronounced coupling to surface plasmons and an increased LDOS have been used in a number of  MNP configurations, such as single photon transistors~\cite{Sorensen:Nature07}, chemical detection/imaging~\cite{Negro09,Hou:13}, optical antennas~\cite{Maitre13},
spasing~\cite{Bergman:PRL03,Naginov:Nature09}, enhanced spontaneous emission~\cite{Novotny06, Maitre13}, and  long range  entanglement~\cite{David10}. 

Accurate quantitative modelling of  dipole emitters near arbitrarily shaped MNPs is challenging.{ In many cases, however, the LDOS enhancements in  cavity structures are directly attributable to one or just a few local resonances which may be rigorously described mathematically as quasinormal modes
(QNMs) of the open system~\cite{EPAWSK98}. In particular, it has been shown that the scattering resonance of the MNPs can be described by QNMs~\cite{deLasson_JOSAB_30_1996_2013,Sauvan13}; so intuitively, it should be possible to capture most of physics of the dipole-MNP interaction around the local plasmon resonances, using just a few QMNs.}
  Indeed, this is the typical approach in dielectric cavity geometries where, e.g., the Purcell factor~\cite{Purcell} for dipole emitters within the cavity can be accurately determined in terms of the cavity mode quality factor ($Q$) and mode volume ($V$). However, as recently shown in Ref.~\cite{Philip:OL12},
{even in this relatively simple situation,  the mode volume is in fact non-trivial to define in a rigorous way because of the leaky nature of the cavity modes which causes the field to diverge far from the cavity. In dealing with MNPs, the situation is {\it further} complicated because the dielectric constant of the MNP is complex and strongly dispersive.}
While some authors believe that it is unnatural to work with modes in a  lossy system~\cite{SipePRA:2012}, {we argue that QNMs have} enormous intuitive appeal in MNP geometries, and {that} they help to accurately explain the underlying physics of light-matter coupling in a remarkably clear and transparent way. {The use of QNMs in the field of nanoplasmonics is made difficult by a number of challenges:} %
($i$) techniques developed in quantum optics for lossy inhomogeneous structures suggest that traditional mode expansion techniques do not work~\cite{Welsch:PRA99}; ($ii$) {proper calculations of localized surface plasmons as QNMs are non-trivial; 
{($iii$) because  QNMs diverge in space, it is impossible to directly use them in calculations of the electromagnetic propagator to positions far away from the resonators; and ($iv$): the LDOS is known to diverge near a metal surface due to quasi-static coupling (e.g., causing Ohmic heating) so at these positions a single QNM expansion is not expected to work.


In this paper, we describe a new QNM expansion technique that can be used to evaluate the electric field from a dipole emitter at positions both nearby and far away  from MNPs {(i.e., outside the resonator)}. 
We first compute the effective mode volumes and Purcell factors for metal nanorods, and compare the resulting approximate results to full numerical calculations. 
We elaborate on the concept of mode volumes for the QNMs of MNPs~\cite{Koen10} and show explicitly that a recently introduced mode volume for MNPs~\cite{Sauvan13} is exactly the generalization of the result in Ref.~\cite{Philip:OL12} to the case of dispersive materials. 
We  present a solution to the problem of using QNM expansions in scattering calculations at positions outside {and far away from the resonator where the modes  diverge~\cite{Leung941}}. Further, we introduce an  analytic technique to account for Ohmic lossesquasi-static coupling which become important when the emitter is within just a few nm of a MNP. 

\section{Green function expansion in terms of  normalized quasinormal modes}

Consider a general, {non-magnetic}, inhomogeneous medium {described by a} 
complex permittivity $\varepsilon({\bf r},\omega)=\varepsilon_\text{R}({\bf r},\omega)+{\rm i}\varepsilon_\text{I}({\bf r},\omega)$. 
{{For a 3D geometry}, the total photon  Green function  satisfies the} equation
\begin{align} 
 \nabla \times \nabla \times {\bf G}({\bf r},{\bf r}';\omega) -k_{0}^{2} \varepsilon({\bf r},\omega) {\bf G}({\bf r},{\bf r}';\omega) = k_{0}^2\delta({\bf r}-{\bf r}'){\bf 1},
\end{align}
where $k_0=\omega/c$ 
and ${\bf 1}$ is the unit dyadic. To   describe quantum optical effects in lossy 
inhomogeneous structures~\cite{Welsch:PRA99,BarnettPRL68,SuttorpJPB2010,yao:PRB09}, two key Green functions are required, namely
${\bf G}({\bf r}_a,{\bf r}_a)$ and ${\bf G}({\bf r}_a,{\bf r}_b)$; the former   
  can describe effects such as modified spontaneous emission (SE) or the Lamb shift \cite{yao:PRB09}, 
while the latter  
includes the  effects of photon propagation. 
All observables in quantum optics, e.g.,   dipole-emitted spectra, {can be described in terms of}  
these two  quantities~\cite{Welsch:PRA99,Vanvlack:PRB12,Hughes13}. For example, if we consider a  dipole emitter at position ${\bf r}_a$ with a dipole  moment  ${\bf d}=d {\bf n}_{a}$ (${\mn}_a$ is a unit vector), the
relative SE emission rate is~\cite{NovotnyAndHecht_2006}
\begin{align}
F_a({\bf r}_a,\omega)=\frac{\Gamma_{a}(\mr_{a},\omega)}{\Gamma_\text{B}(\omega)} = \frac{\text{Im}\left\{{\mn}_a\cdot{\bf G}({\bf r}_{a},\mr_a;\omega)\cdot{\mn}_a\right\}}{\text{Im}
\left\{{\mn}_a\cdot\mG_\text{B}
(\mr_a,\mr_a;\omega)
\cdot{\mn}_a\right\}},
\label{Eq:PurcellFromLDOS}
\end{align}
where ${\bf G}_{\rm B}$  is the    homogeneous medium  Green function with $\varepsilon(\mr)=\varepsilon_\text{B}=n_\text{B}^2$ 
\cite{NovotnyAndHecht_2006}.
{The Green functions may be computed in a number of ways~\cite{Welsch:PRA99,YaoLaserReviews}, but in general they are rather expensive to calculate. 

Localized surface plasmons may be directly understood as QNMs of the MNPs \cite{deLasson_JOSAB_30_1996_2013}, defined as the frequency domain solutions to the wave equation with open boundary conditions~\cite{Lee99, Leung941} (the Silver-M\"uller radiation condition~\cite{Martin_MultipleScattering}). {This makes QNMs the natural starting point for theoretical developments, although the radiation condition is not immediately compatible with typical mode solvers. For this reason, a common approach is 
the use of coordinate transforms, typically in the form of Perfectly Matched Layers (PMLs), to model a system with no reflections from the simulation domain boundaries. The QNMs, $\mft_\mu(\mr)$, have a discrete spectrum of complex resonance frequencies, $\tilde{\omega}_{\mu}=\omega_\mu-{\rm i}\gamma_\mu$, from which the resonator quality factor is 
$Q=\omega_\mu/2\gamma_\mu$. An important consequence of the complex resonance frequency 
 is that the QNMs diverge  (exponentially) in the limit $r\rightarrow\infty$. 


In certain spatial regions,  {such as { inside} the resonator~\cite{Leung941}},
the transverse part of the Green function can be expanded as~\cite{Lee99}
${\bf G}^{\rm T}({\bf r}_1,{\bf r}_2;\omega)= \sum_{\mu}\frac{\omega^2}{2\tilde\omega_{\mu}(\tilde\omega_{\mu}-\omega)}\tilde{\bf f}_{\mu}({\bf r}_1)\tilde{\bf f}_{\mu}({\bf r}_2)$.
{This has been rigorously proven for spheres, but is also expected to be true for non-spherical scattering objects that have an abrupt discontinuity in the dielectric constant profile \cite{Lee99}}.  Thus for a single QNM,
considering points inside the resonator, we define
\begin{align}
{\bf G}^{\rm f}_\mu({\bf r}_1,{\bf r}_2;\omega) \equiv \frac{\omega^2}{2\tilde\omega_{\mu}(\tilde\omega_{\mu}-\omega)}\tilde{\bf f}_{\mu}({\bf r}_1)\tilde{\bf f}_{\mu}({\bf r}_2),  
\label{eq:1}
\end{align}
where ${\bf f}_{\mu}$ is normalized through~\cite{Leung941}
 \begin{align}
\langle\langle \tilde{\bf f}_{\mu}|\tilde{\bf f}_{\nu}\rangle\rangle\!&=\!\lim_{V\rightarrow\infty}\!\int_V\left(\frac{1}{2\omega}\frac{\partial (\varepsilon({\bf r},\omega)\omega^2)}{\partial \omega}\right)_{\omega=\tilde{\omega}_{\mu}}\!\!\!\tilde{\bf f}_{\mu}({\bf r})\!\cdot\!\tilde{\bf f}_{\nu}({\bf r})d{\bf r} 
\nonumber\\
&
+ i\frac{n_\text{B}c}{2\tilde{\omega}_{\mu}}\int_{\partial V}\tilde{\bf f}_{\mu}({\bf r})\cdot\tilde{\bf f}_{\nu}({\bf r})d{\bf r}=\delta_{\mu\nu}.
\label{eq:norm}
\end{align}

\begin{figure}[b]
\centering\includegraphics[trim=0.cm 0.cm 0.cm 0.cm, clip=true,width=.8\columnwidth]{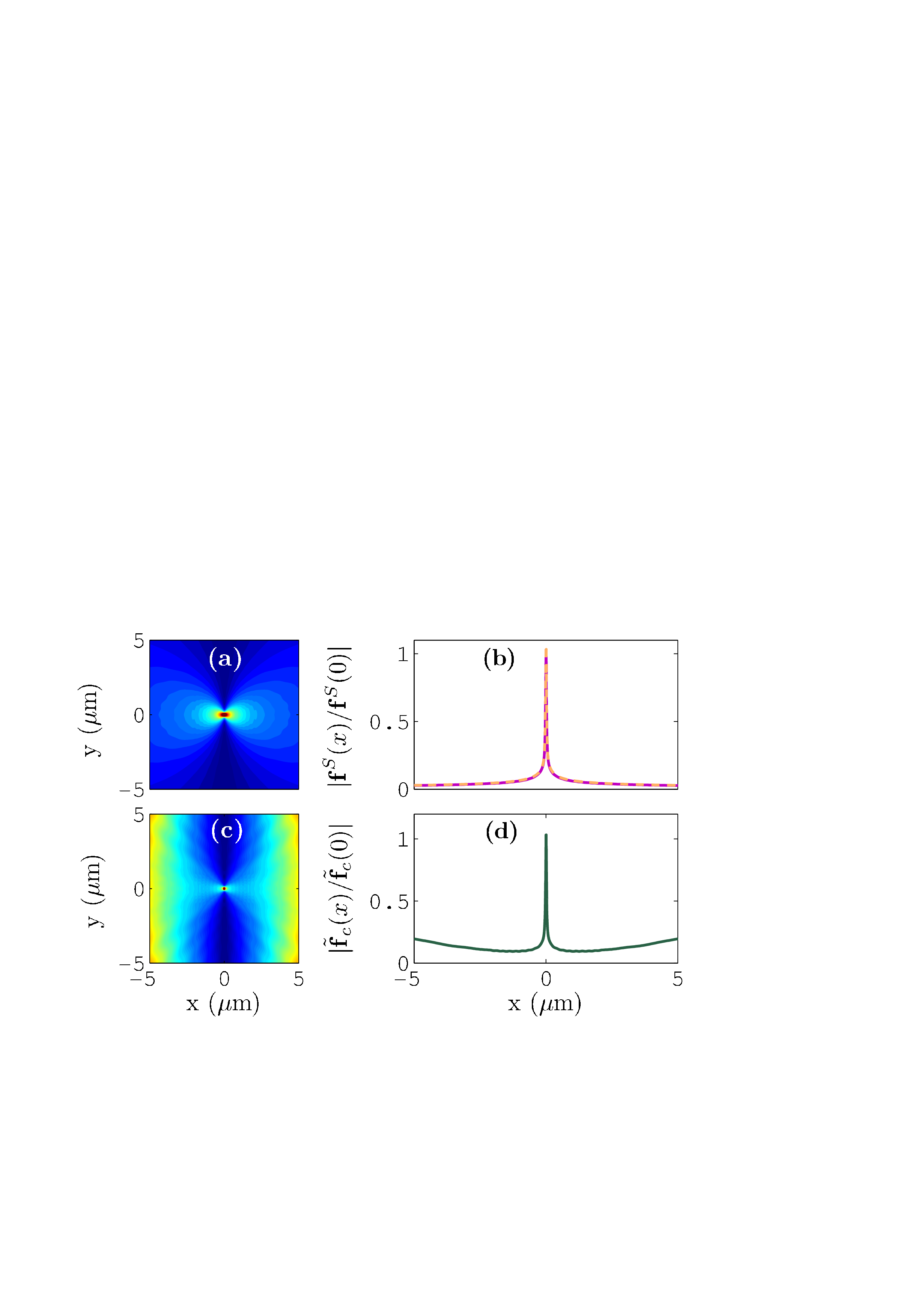}
\caption{Computed fields for the 2D MNP rod. Spatial profiles of the electric field calculated at $\omega_c={\rm Re}(\tilde\omega_c)$, where  $\tilde\omega_{\rm c}/2\pi=
415.863 - {\rm i}37.176$~THz ($1.720 - {\rm i}0.154$ eV). 
(a-b): Normalized  scattered field $|{\bf f}^{\rm S}(\omega_{\rm c})|$ (solid) and  regularized QNM field  $|\tilde{\bf F}(\omega_{\rm c})|$ (dashed). 
(c-d)  QNM $|\mft_\text{c}(\omega_{\rm c})|$.  
}
\label{fig:1}
\end{figure}

As a representative  example, we   first study a 2D MNP using a
Drude model, $\varepsilon(\omega)=1-\frac{\omega_{\rm p}^2}{\omega(\omega+i\gamma)}$.
 We  consider a metal rod with  width ($x$ axis) $10$~nm, and length ($y$ axis) $80$~nm  located in a homogeneous space with $n_{\rm B}=1.5$,  where  $\omega_{\rm p} = 1.26\times10^{16}~{\rm rad/s}$ and $\gamma=7\times10^{13}~{\rm rad/s}$. To compute the QNM profile, $\tilde{\bf f}_{\rm c}({\bf r})$, we use the FDTD technique~\cite{FDTD} with PML boundary conditions {and a run-time Fourier transform}. 
Specifically, the system is excited using a pulsed plane wave (incident along $x$, and polarized in $y$, which was chosen to maximally excite the QNM of interest) and a temporal window function is applied for computing the field.
{Figure~\ref{fig:1} shows the spatial profiles of the QNM field as well as the {\em scattered} part of the field evaluated at the real part of the complex eigenfrequency. 
Only the QNM shows the expected spatial divergence~\cite{Philip:OL12}}. 
Later we will introduce
a regularized QNM field that  coincides very well with  the
scattered field profile, as shown in Fig.~\ref{fig:1}(d) (dashed curve);
this regularization is essential for modelling light-matter interactions outside the MNP. 

\section{Effective mode volume and Purcell factors for
spatial locations beyond the quasi-static coupling regime}

 For most problems in light-matter interactions, one  must 
  describe how light behaves far away from the MNP. Examples include  calculations of the  spectrum that would be detected at some
 arbitrary spatial location~\cite{Welsch:PRA99,Vanvlack:PRB12,Hughes13},
 and in the design of nanoantennas \cite{NovotnyAndHecht_2006}. {The QNMs diverge at large distances, and therefore it is impossible to use them directly for such calculations.}  
However, since we do have a  {highly accurate approximation} for the Green function {\em inside} the resonator, we can exploit a Dyson equation of the form ${\bf G} = {\bf G}^{\rm B} +  {\bf G}^\text{B} \cdot\Delta\varepsilon {\bf G}$ ($\Delta\varepsilon=\varepsilon_{\rm MNP}-\varepsilon_{\rm B}$)   to obtain a corrected Green function {in the region  far away from} the scattering geometry (see \ref{AppB}). {With this approach, we derive a corrected Green function expression for positions {\em outside} the resonator,
\begin{align}
{\bf G}^{\rm far}({\bf r}_1,{\bf r}_2;\omega) =
{\bf G}^{\rm B}({\bf r}_1,{\bf r}_2;\omega)+{\bf G}^{\rm F}_\mu({\bf r}_1,{\bf r}_2;\omega),\\
  \label{eq:Gout}
 {\bf G}^{\rm F}_\mu({\bf r}_1,{\bf r}_2;\omega) =
\frac{\omega^2}{2\tilde\omega_{\mu}
(\tilde\omega_{\mu}-\omega)}\tilde{\bf F}_{\mu}({\bf r}_1)\tilde{\bf F}_{\mu}({\bf r}_2),
\end{align}
 where we have introduced a new {regularized} field, 
 \begin{align}
\tilde{\bf F}_{\mu}({\bf r})\equiv \int_{V}{\bf G}^{\rm B}({\bf r},{\bf r'})\cdot\Delta\varepsilon({\bf r'})\tilde{\bf f}_{\mu}({\bf r}')d{\bf r}'.
\end{align}
As shown in Fig.~\ref{fig:1}(b), this field coincides well with the scattered field
because it is essentially the field that is scattered by the QNM at the
(real) frequency $\omega_c$.
In Eq.~(\ref{eq:Gout}), we have neglected the influence of other  mode contributions   
 as we are currently interested in positions sufficiently far from the MNP, that these can be safely ignored.}

\begin{figure}[b]
\centering\includegraphics[trim=0.cm 0.cm 0.cm 0.cm, clip=true,width=.8\columnwidth]{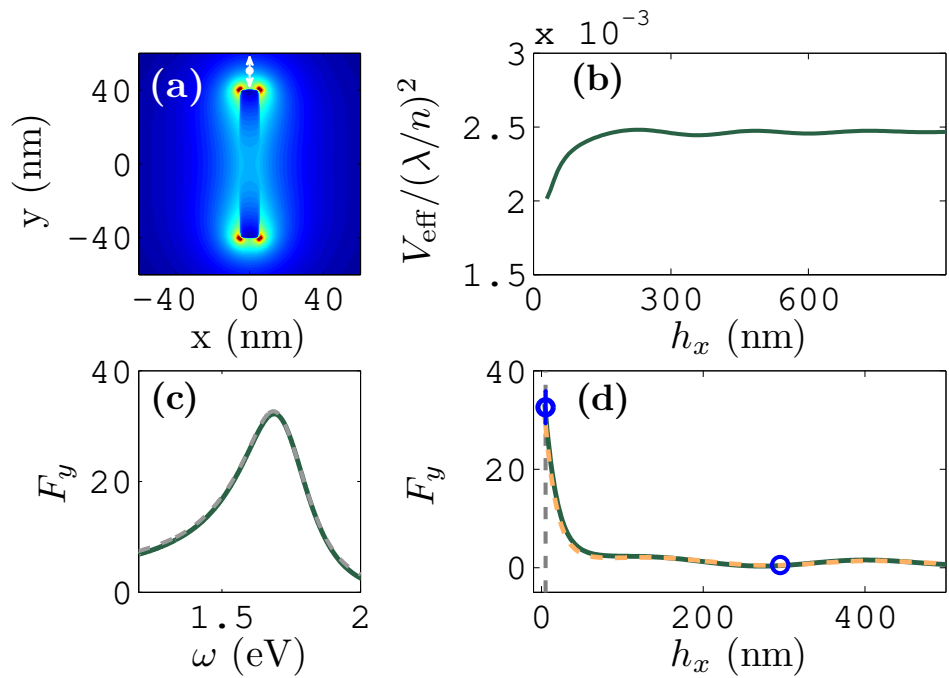}
\caption{{\em }  (a) Expanded view of  QNM profile, $|\mft_\text{c}|$, from Fig.~\ref{fig:1}(c).
(b) Computed QNM effective mode volume $V_{\rm eff}$ as a function of the distance from the MNP surface {to the domain boundary}. 
(c) Enhanced SE factor, $F_y({\bf r}_y,\omega)$ at {${\bf r}_y$ = (0,$10.4~$nm)} above the top side of the MN (see arrow in (a)) using ${\bf G}^{\rm f}$ (green/dark solid), and dipole solution (grey/light dashed). (d) On-resonance  SE factor, $F_y(\omega_c)$, as a function of distance spanning 5~nm (vertical dashed curve) to   500~nm away from the top 
edge of the MNP (orange dashed), and with ${\bf G}^{\rm f}$ in place of ${\bf G}^{\rm F}$ (green solid); the circles indicate the full dipole results at selected locations. }
\label{fig:2}
\end{figure}

{We now consider an oscillating  dipole with (real) resonant frequency near the (complex) frequency of the QNM in Fig.~\ref{fig:1}, so that a single mode expansion ($\mu=\text{c}$) of the Green function might be expected to provide a good approximation, at least for dipole-MNP separations beyond the {quasi-}static coupling regime.}
The next higher order plasmon mode is at least 500 meV, so we can safeuly
assume a single mode approximation. 
 To describe the relative SE rate in terms of the Purcell factor, 
we write
\begin{align}
F_a({\bf r}_a,\omega) = F_\text{P}\,\eta({\mr}_a,{\mn}_a;\omega)+1,
\label{Eq:GeneralizedPurcell}
\end{align}
where the Purcell factor {has} the familiar form \cite{note2D}
\begin{align}
F_\text{P} = \frac{3}{4\pi^2}\left(\frac{\lambda}{n_{\rm B}}\right)^3 \left (\frac{Q}{V_{\rm eff}} \right),
\label{eq:PF}
\end{align}
and the factor $\eta({\mr}_a,{\mn}_a;\omega)$ \cite{Gerard_JLT_17_2089_1999} %
accounts for  deviations of the emitter at ${\bf r}_a$ from the field maximum $\mr_0$, cavity polarization, and cavity resonant frequency {{(see \ref{AppC})}. 
Starting from Eqs.~(\ref{Eq:PurcellFromLDOS})-(\ref{eq:1}), in the limit $\gamma_c \ll \omega_c$, we derive the  QNM effective mode volume  
\begin{align}
\frac{1}{V_{\rm eff}}  = \text{Re}\left\{\frac{1}{v_\text{Q}} \right\},\quad v_\text{Q} = \frac{\langle\langle\mft_\text{c}|\mft_\text{c}\rangle\rangle}{\varepsilon_{\rm B}\mft_\text{c}^2({\bf r}_0)} ,
\label{eq:Vq}
\end{align}
in agreement with previous generalized mode volume results for a lossless dielectric cavity~\cite{Philip:OL12}.
We note that Sauvan \emph{et al.} \cite{Sauvan13} recently used the Lorentz reciprocity theorem to derive
 an expression for the effective mode volume which 
can be shown to be identical to Eq.~(\ref{eq:Vq}) (see \ref{AppA}). Figures~\ref{fig:2}(a) and \ref{fig:2}(b) show, respectively, the QNM spatial profile in more detail and the evaluation of $V_{\rm eff}$ 
as the size of the calculation domain is increased.
 Although each  term in Eq.~(\ref{eq:norm}) diverges as a function of domain size, the sum  converges quickly to a finite value
after  $\sim$500~nm. 
In optical fiber geometries, this crossover region is referred to as the ``caustic radius'' (${\bf r}_{\rm caustic}$)~\cite{SynderLoveBook}.
Although one can define the SE rate 
in terms of the Purcell factor as in Eq.~(\ref{Eq:GeneralizedPurcell}), it may in practice be more convenient to work directly with Eqs.~(\ref{Eq:PurcellFromLDOS}) and (\ref{eq:Gout}) as we do below.


%
\begin{figure}[b]
\centering\includegraphics[trim=0.cm 0.cm 0.cm 0.cm, clip=true,width=.8\columnwidth]{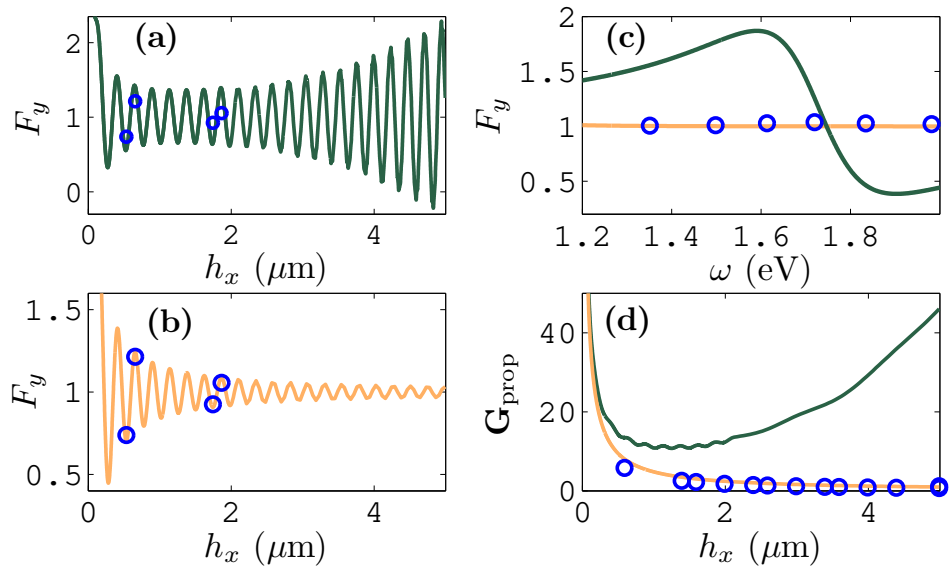}
\caption{Enhanced SE factor and propagator for a 2D MNP, where circles indicate full dipole results. (a) 
 On-resonance SE factor 
as a function of ${\bf r}_a=(h_x,0)~$ (with $h_x>100~$nm) 
using ${\bf G}^{\rm f}$ in place of ${\bf G}^{\rm F}$
in Eq.~(\ref{eq:Gout}). (b) As in (a), but using ${\bf G}^{\rm F}$. (c) SE factor as a function of frequency at  ${\bf r}=(5,0)~\mu$m
(orange/light), and with ${\bf G}^{\rm f}$ in 
place of ${\bf G}^{\rm F}$ (green/dark solid). (d) Propagator $|{\bf G}_{yy}({\bf r}_{a},{\bf r}_b;\omega_c)/{\rm Im}[{\bf G}^{\rm B}_{yy}({\bf r}_a,{\bf r}_a;\omega_c)]|^2$ using ${\bf G}^{\rm f}$ (green/dark solid) and ${\bf G}^{\rm F}$ (orange/light solid), as a function of  ${\bf r}_b$ (with ${\bf r}_a=(10,0)~$nm).}
\label{fig:3}
\end{figure}

Figure~\ref{fig:2}(c) shows the relative SE rate {as a function of frequency} for a dipole aligned along the {nanorod} direction and located {just above} 
the top interface. The grey dashed line is the result from {a full, independent numerical calculation using FDTD~\cite{Yao10,FDTD}, i.e., with no mode expansions
(namely, a full dipole calculation), 
and the green solid is the result from {Eq.~(\ref{Eq:GeneralizedPurcell})}. Clearly, the  analytic  SE rate gives a very good  fit to the full dipole calculation at this spatial location, including the entire non-Lorentzian lineshape. 
Figure~\ref{fig:2}(d) depicts the on-resonance $F_a(\omega_c)$, as a function of distance (5~nm to 500~nm) from the MNP. {For these distances, the single
QNM expansion in Eq.~(\ref{eq:Gout}) 
 provides an excellent approximation to the full dipole results,  even if one uses $\tilde{\bf f}_{\mu}$ in place of $\tilde{\bf F}_{\mu}$   (see Figs.~\ref{fig:1}(a) and \ref{fig:1}(d) over this range).  However, the approximation must  fail  at evaluating the SE rate for larger emitter-MNP separations, and for evaluating the propagated fields, since the rigorous 
 QNMs, $\tilde{\bf f}_\mu$,  diverge in space.
In Fig.~\ref{fig:3}(a) we show the on-resonance SE factor as a function of distance, spanning 5~nm to 5~$\mu$m using ${\bf G}^{\rm f}$ in place of ${\bf G}^{\rm F}$.
Comparing to the full dipole solution, the single QNM approximation clearly fails to get the correct far field behavior, as anticipated from our  discussions above. In contrast, the SE factor computed using ${\bf G}^{\rm F}$ in ${\bf G}^{\rm far}$ (see Fig.~\ref{fig:3}(b)) shows excellent agreement with the full dipole results and yields the correct behavior for large distances {where it tends to unity.} 
 Importantly, in both cases we only include $\mu={\rm c}$ in the Green function summation, although ${\bf G}^{\rm F}$ accounts for the influence {of light propagation}. 
In  Fig.~\ref{fig:3}(c) we show the two model results as a function of frequency at ${\bf r}=(5,0)~\mu$m, and  ${\bf G}^{\rm f}$ again yields  incorrect  results.  In Fig.~\ref{fig:3}(d), we  show the {absolute square of} the  propagator, 
$|{\bf G}_{yy}({\bf r}_{\rm a},{\bf r}_b;\omega_c)|^{2}$, 
as a function of  ${\bf r}_b$,
with selected full-dipole results shown with the circles; once again, we observe that only ${\bf G}^{\rm F}$ gives the correct behavior at large distances.



\section{Enhanced spontaneous emission factors for
spatial locations very near the MNP resonator}

Together, Figs.~\ref{fig:2}(d) and Fig.~\ref{fig:3}  demonstrate that the
regularized QNM fields  $\tilde{\bf F}_{\mu}$, through ${\bf G}^{\rm far}$, provide an accurate representation of ${\bf G}$ for all separations down to at least $\sim5~$nm.  However, ultimately ${\bf G}^{\rm far}$ must fail at even shorter dipole-MNP separations
because of the known {quasi-static} divergence of the LDOS. 
This divergent behavior 
cannot be accurately} accounted for in a single QNM approximation. {Our solution to this problem is to note that the dynamics of} a dipole near a MNP surface {is governed by} essentially the same LDOS increase as a dipole near a metal half space. {Moreover, this divergent term vastly dominates the LDOS}, and therefore, 
at spatial points very near the resonator, we {may} 
simply add on this known quasi-static Green function to the single QNM approximation to the Green function, so that 
\begin{align}
{\bf G}^{\rm out}({\bf r}_a,{\bf r}_b;\omega) =
{\bf G}^{\rm far}({\bf r}_a,{\bf r}_b;\omega) + {\bf G}^{\rm qs}({\bf r}_a,{\bf r}_b;\omega),
\label{eq:Ges}
\end{align}
where the quasi-static term is given by \cite{MartinES2001}
$
{\bf G}^{\rm qs}({\bf r}_a,{\bf r}_b) = 
\mp {\bf G}^{\rm B}({\bf r}_a,-{\bf r}_b)
\frac{\varepsilon(\omega)-\varepsilon_{\rm B}}{2(\varepsilon(\omega)+\varepsilon_{\rm B})}$, {and} $\mp$ is for $s$- or $p$-polarized dipoles, respectively.
We find this formula to be quantitatively correct for all positions that we have tried.
For further details, see  \ref{AppB}.

\begin{figure}[h]
\centering\includegraphics[trim=0.cm 0.cm 0.cm 0.cm, clip=true,width=.8\columnwidth]{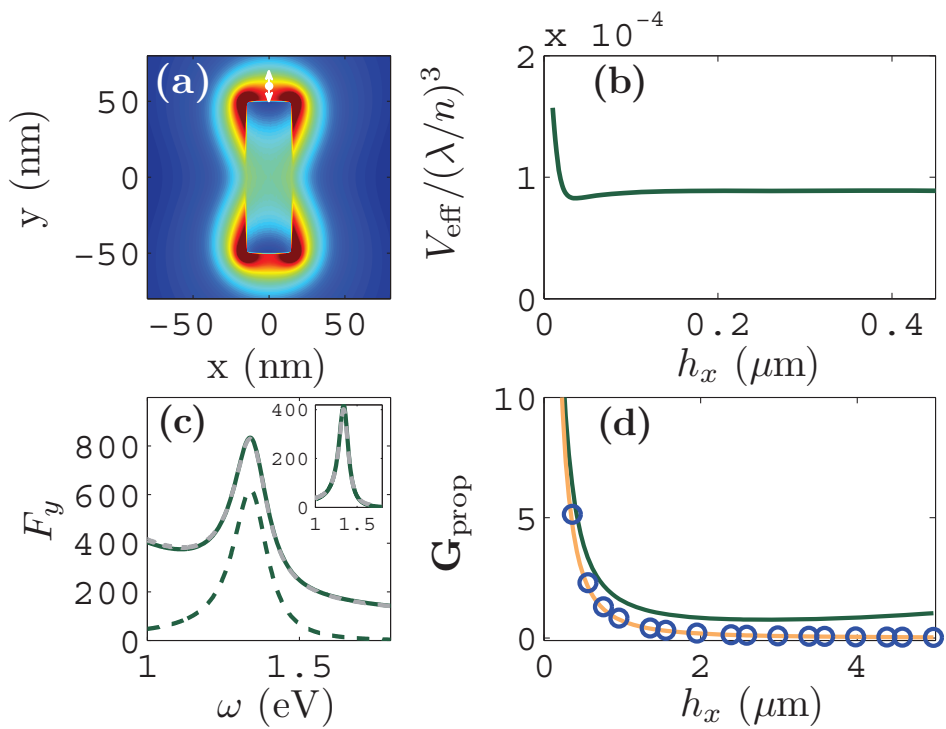}
\caption{{Mode volume and enhanced SE factor for a 3D gold nanorod.} (a)  QNM profile, $|\tilde{\bf f}(x,y,0;\omega_{\rm c})|$.
(b) Effective mode volume, $V_{\rm eff}$  as a function of the distance from the MNP surface {to the domain boundary}. (c) SE factor, $F_y({\bf r}_y,\omega)$ at 2~nm above the surface of the MNP (see arrow in (a)) using ${\bf G}^{\rm f}$ in place of ${\bf G}^{\rm F}$ with (green solid) and without (green/dark dashed) quasi-static correction, and full dipole solution (grey/light dashed). Inset shows the same calculation at 10~nm with no quasi-static correction. (d) Propagator $|{\bf G}_{yy}({\bf r}_{a},{\bf r}_b;\omega_c)/{\rm Im}[{\bf G}^{\rm B}_{yy}({\bf r}_a,{\bf r}_a;\omega_c)]|^2$ 
(orange/light solid) as a function of  ${\bf r}_b$ (with ${\bf r}_a=(10,0)~$nm), and  using ${\bf G}^{\rm f}$ in place of ${\bf G}^{\rm F}$  (green/dark solid); circles indicate  full dipole results.  }
\label{fig:4}
\end{figure}

To demonstrate the accuracy of Eq.~(\ref{eq:Ges}), we
 now  consider a 3D metal nanorod {(i.e., a cylinder)} with
radius $15$~nm and length $100$~nm, where the rotational axis coincides with the $y$ axis (see Fig.~\ref{fig:4}(a)). The Drude parameters are the same as before, but we  use $\gamma=1.41\times10^{14}~{\rm rad/s}$ to model gold. We find a frequency region dominated by a single QNM at $\tilde{\omega}_{\rm c}/(2\pi) =
 324.981-{\rm i}16.584~$THz (1.344 - {\rm i}0.0684 eV), and the next high-order mode is at least 600 meV away. 
The  mode profile is shown in Fig.~\ref{fig:4}(a), and  $V_{\rm eff}$   is shown in Fig.~\ref{fig:4}(b). 
In Fig.~\ref{fig:4}(c) we show
  the SE spectrum for a dipole emitter  at 2~nm (see arrow in Fig.~\ref{fig:4}(a)). The green dashed  curve uses Eq.~(\ref{eq:Gout}) (with ${\bf G}^{\rm f}$ in place of ${\bf G}^{\rm F}$),
while the grey dashed curve is the full dipole result which is seen to be considerably different; in contrast, the proposed ${\bf G}^{\rm out}$ model (solid green) is seen to be in quantitative agreement with the full dipole result even in this spatial regime dominated by non-radiative coupling.
In Fig.~\ref{fig:4}(d),
we also show the propagator and confirm an excellent fit as a function of distance {\em only} if ${\bf G}^{\rm F}$ is used.

\section{Conclusions}

We have introduced a   QNM expansion technique for arbitrarily shaped  MNPs and 
numerically demonstrated its accuracy in evaluating the enhanced SE rate and far-field propagator associated with proximate dipole oscillators.
In the limit of a single QNM resonance and spatial positions outside the regime of quasi-static coupling, we give a closed form solution for the Purcell factor. The corresponding effective mode volume is found to be identical to the generalized effective mode volume recently introduced
for dielectric cavities~\cite{Philip:OL12}, with a slight generalization to account for material dispersion.
  Our results are obtained using a transparent  analytic expression for the photon Green function  evaluated outside of the MNP, 
 ${\bf G}^{\rm out}= {\bf G}^{\rm B}+  {\bf G}^{\rm F} +{\bf G}^{\rm qs}$,
that is valid everywhere outside the particle's boundary.  The ${\bf G}^{\rm qs}$ term is a simple, intuitive quasi-static dipole expression that accounts for quasi-static coupling within a few nanometers of the metal resonator, while ${\bf G}^{\rm far}={\bf G}^{\rm B} + {\bf G}^{\rm F}$  is the sum of the familiar Green function of a homogenous medium, and a resonant mode expansion.  
Our  general approach  can be applied to a wide variety of  resonator geometries including
hybrid metal--photonic-crystal structures~\cite{Barth}, 
which are otherwise very hard  to understand and model.


\section*{Acknowledgements}
This work was supported by the Natural Sciences and Engineering Research Council of Canada, the Canadian Institute for Advanced Research, Queen's University {and the Danish Council for Independent Research (FTP 10-093651)}.
We thank Cole Van Vlack for useful discussions.

\appendix

\section{Dyson equation approach for obtaining the  Green function and regularized quasinormal mode}
\label{AppB}

{It was shown in Ref.~\cite{Leung941} for a one-dimensional lossy cavity that if the permittivity or any order of its derivative is discontinuous at the border of the cavity, then the QNMs inside the cavity form a complete basis; so one can use a  mode expansion formulation for the Green function in terms of the QNMs. Later, Ref.~\cite{Lee99} proved the same result  for a three-dimensional sphere, and they also discussed why the same argument could be applied to structures without spherical symmetry. Bergman and Stroud~\cite{DJB80} used the same argument for scattering geometries made up of spherical structures. For our calculation of nanorods, we also find that a single QNM expansion works exceptionally  well, and, moreover, we formally  recover the same result as in Ref.~\cite{Sauvan13} (which uses an entirely different method). Consequently we can assume that the mode expansion formulation approach is valid for points inside resonators of any shape of geometry as long as the permittivity (or its derivative) is discontinuous at the border of the resonator.}
To obtain the {corrected} Green function for points {\em outside} the scattering geometry (MNP), we utilize the  Dyson equation,
 \begin{align}
 {\bf G}({\bf r}_1,{\bf r}_2;\omega)= 
 {\bf G}^{\rm B}({\bf r}_1,{\bf r}_2;\omega)+
 \int_{V}{\bf G}^{\rm B}({\bf r}_1,{\bf r'}_1;\omega)\cdot\Delta\varepsilon({\bf r'}_1,\omega){\bf G}({\bf r'}_1,{\bf r}_2;\omega)d{\bf r'}_1,
 \label{eq:1_App}
 \end{align}
 where $\Delta\varepsilon({\bf r},\omega) =\varepsilon_{\rm MNP}({\bf r},\omega)- \varepsilon_{\rm B}$, with $\varepsilon_{\rm B}$ and ${\bf G}^{\rm B}$ the dielectric constant and Green function of the homogeneous background {in which} 
the MNP is embedded.{ For ${\bf r}_1'$ inside the scattering geometry
\begin{align}
{\bf G}({\bf r}_2,{\bf r}_1';\omega) = {\bf G}^{\rm B}({\bf r}_2,{\bf r}_1';\omega) + \int_V{\bf G}^{\rm B}({\bf r}_2,{\bf r}_2';\omega)\cdot\Delta\varepsilon({\bf r}_2',\omega){\bf G}({\bf r}_2',{\bf r}_1';\omega)d{\bf r}_2'.
\label{new1_App}
\end{align}
Substituting the QNM expansion of ${\bf G}({\bf r}_2',{\bf r}_1';\omega)$, for which both points ${\bf r}_2'$ and ${\bf r}_1'$ are inside the MNP, into Eq.~(\ref{new1_App}), and focusing on the single QNM of interest, we obtain
\begin{align}
{\bf G({\bf r}_2,{\bf r}_1';\omega)} = {\bf G}^{\rm B}({\bf r}_2,{\bf r}_1';\omega) + \frac{\omega^2}{2\tilde{\omega}_\mu(\tilde{\omega}_\mu-\omega)}\int_V{\bf G}^{\rm B}({\bf r}_2,{\bf r}_2';\omega)\cdot\Delta\varepsilon({\bf r}_2',\omega)\tilde{\bf f}_\mu({\bf r}_2')\tilde{\bf f}_\mu({\bf r}_1')d{\bf r}_2'.
\label{new2_App}
\end{align}
By inserting the transpose of Eq.~(\ref{new2_App}) into Eq.~(\ref{eq:1_App})},
we derive  the following  Green function      
\begin{align}
{\bf G}^\rev{\rm }({\bf r}_1,{\bf r}_2;\omega)&{=}
{\bf G}^{\rm B}({\bf r}_1,{\bf r}_2;\omega)
+ \frac{\omega^2}{2\tilde\omega_\mu(\tilde\omega_\mu-\omega)}\tilde{\bf F}_\mu({\bf r}_1,\omega)\tilde{\bf F}_\mu({\bf r}_2,\omega) + \mG^\text{others}(\mr_1,\mr_2,\omega),
\label{eq:2_App}
\end{align}
{in which ${\bf r}_1$ and ${\bf r}_2$ are{} two space points {\em outside} the resonator and} we have introduced a new mode field, defined as 
\begin{align}
\tilde{\bf F}_\mu({\bf r},\omega)\equiv \int_{V}{\bf G}^{\rm B}({\bf r},{\bf r'};\omega)\cdot\Delta\varepsilon({\bf r'},\omega)\tilde{\bf f}_\mu({\bf r}')d{\bf r}'.
\label{eq:3_App}
\end{align}
{In Eq.~(\ref{eq:2_App}), $\mG^\text{others}$ formally includes the contributions from all {\em other} QNMs except $\tilde{\bf F}_\mu(\mr)$ which is typically by far the most dominant field at distances far enough away from the MNP that quasi-static effects may be neglected.} 
As we show in the main text, {Eq.~(\ref{eq:2_App}) provides an excellent approximation to $\mG$ at most points outside the resonator
resonator, when the $\mG^\text{others}$ term is neglected, appart from very close to the metal surface where it only fails for $x\leq$ 5 nm.}.

For convenience, we define a new modal Green function in terms of the  new mode field,
\begin{align}
{\bf G}^{\rm F}_\mu({\bf r}_1,{\bf r}_2;\omega) = \frac{\omega^2}{2\tilde\omega_\mu(\tilde\omega_\mu-\omega)}\tilde{\bf F}({\bf r}_1,\omega)\tilde{\bf F}({\bf r}_2,\omega),
\label{eq:4_App}
\end{align}
{in which case we can write the approximate Green function at positions where we can safely neglect $\mG^\text{others}$ as $\mG^\text{far}=\mG^\text{B}+\mG^{\rm F}$. This} can be compared with the single QNM Green function,
\begin{align}
{\bf G}^{\rm f}_\mu({\bf r}_1,{\bf r}_2;\omega)= \frac{\omega^2}{2\tilde\omega_\mu(\tilde\omega_\mu-\omega)}\tilde{\bf f}({\bf r}_1)\tilde{\bf f}({\bf r}_2).
\label{eq:5_App}
\end{align}
These two Green functions (Eqs.~(\ref{eq:4_App})-(\ref{eq:5_App})),  which are both transverse, are used in our manuscript; and  we  demonstrate that only ${\bf G}^{\rm F}$ gets the correct far-field behaviour outside the effective mode volume  region of the QNM ($\geq \sim 500~$nm), while ${\bf G}^{\rm f}$ in place of ${\bf G}^{\rm F}$ yields divergent propagators and divergent {or/and negative} enhanced emission rates for oscillating dipoles {in this limit}. The additional term in Eq.~(\ref{eq:2_App}), namely ${\bf G}^{\rm B}$, also explains exactly {why the extra factor of unity is required to properly relate} 
the relative spontaneous emission (SE) rate {to the Purcell factor}; interestingly,  we note that such a factor does not appear for spatial regimes {\em inside} the scattering geometry, such as with emitters inside photonic crystal cavities {for which Eq.~(\ref{eq:5_App}) is the appropriate single-mode approximation that can be used to derive the Purcell factor}~\cite{Philip:OL12}.

We stress that the only approximation 
is to use the QNM {transverse} Green function {expansion} 
{within} the MNP, with a Dyson equation theory that obtains the correct {transverse} Green function solution outside the scattering geometry.

Neither {$\mG^\text{far}$ or $\mG^\text{f}$} 
are suitable for the extreme near field regime (i.e., a few nm from the MNP surface);
however,   the dominant response of a dipole near a metal surface is very similar to the behaviour of a dipole near a metal half space, and this exact  {\em quasi-static} response is known analytically, through 
 \cite{MartinES2001}
\begin{align}
{\bf G}^{\rm qs}({\bf r}_1,{\bf r}_2;\omega) = 
\mp {\bf G}^{\rm B}({\bf r}_1,-{\bf r}_2;\omega)\,
\frac{\varepsilon_{\rm MNP}(\omega)-\varepsilon_{\rm B}}{2(\varepsilon_{\rm MNP}(\omega)+\varepsilon_{\rm B})},
\label{eq:7_App}
\end{align}
where  $\mp$ is for $s$-polarized or $p$-polarized dipoles, respectively. {Moreover, this term vastly dominates the response at these distances, and therefore we propose to simply add this term in place of $\mG^\text{others}$. Thus, in total we arrive at the approximate analytical Green function for use at {all} positions outside the MNP:
\begin{align}
\mG^\text{out}(\mr_1,\mr_2;\omega) = \mG^\text{B}(\mr_1,\mr_2;\omega) + \mG^{\rm F}(\mr_1,\mr_2;\omega) + \mG^{\rm qs}(\mr_1,\mr_2;\omega).
\end{align}
This expression is both simple and physically appealing in how it describes the various physical processes that occur beyond a single QNM expansion; yet, all that is required is a single QNM supplemented by the normalization procedure for a generalized effective mode volume and quasi-static coupling. In fact our formalism also makes it clear how to simplify complicated numerical calculations requiring large memory and small grids outside the MNP. Since all that is required is the QNM within the MNP, then the  Green function and 
QNM mode can be computed to any arbitrary spatial point outside the MNP analytically. In Fig.~\ref{fig:A}, we show the predicted  SE factor for a quantum dipole {emitter} orientated along the 3D MNP (gold nanorod with a Drude permittivity model---see main manuscript for more details)
using the various Green function models.
In Fig.~\ref{fig:A}(a) we show that calculations using 
${\bf G}^{\rm f}$ (in place of ${\bf G}^{\rm F}$
in ${\bf G}^{\rm far}$)  produce essentially 
identical results  for the distances considered; and since
${\bf G}^{\rm f}$ is easier to compute, one can safely use
${\bf G}^{\rm f}$ for distances less than several hundred nm (see Fig.~\ref{fig:2}(d)). However, as we explicitly in Fig.~\ref{fig:3}(a), for distances outside the {caustic regime,} 
${\bf r}\geq \sim 500~$nm, then the SE rates predicted by ${\bf G}^{\rm f}$
in place of ${\bf G}^{\rm F}$  grow exponentially (which is clearly unphysical), while ${\bf G}^{\rm far}$ consistently produces results that are in very good agreement with full dipole calculations. 
\begin{figure}[t]
\centering\includegraphics[trim=0.5cm 0.8cm 1.cm 1.6cm, clip=true,width=.8\columnwidth]{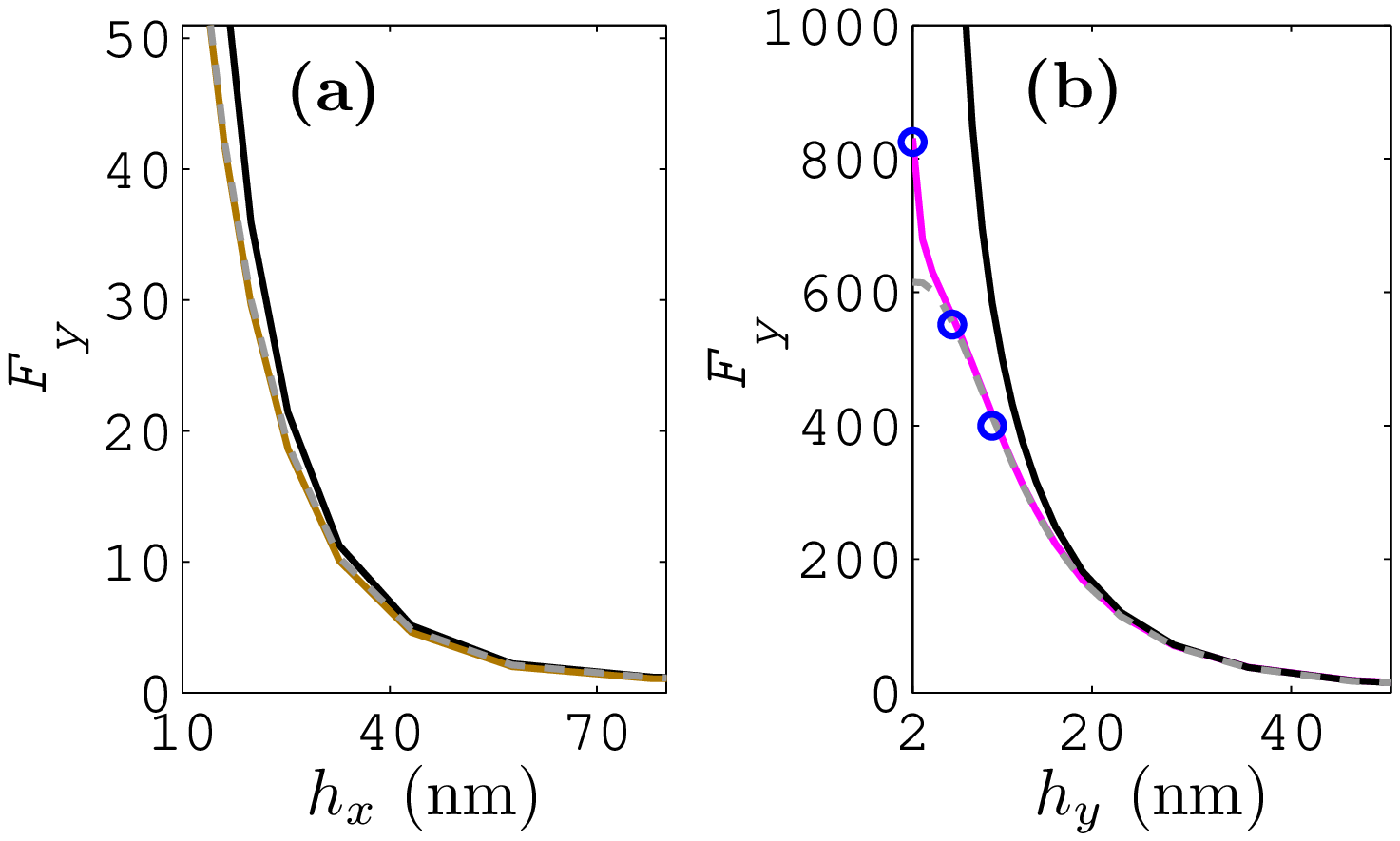}
\caption{{Enhanced SE factor for a 3D gold nanorod.} (a)  SE emission factor as a function of distance from the side of the MNP rod using a $y$-polarized dipole; orange solid (using ${\bf G}^{\rm far}$),
grey dashed (using ${\bf G}^{\rm f}$ in place of ${\bf G}^{\rm F}$), and black solid (using ${\bf G}^{\rm far}+{\bf G}^{\rm back}_{1}$).
(b) SE emission factor as a function of distance from the top of the MNP rod using a $y$-polarized dipole; magenta solid (using ${\bf G}^{\rm B}+{\bf G}^{\rm f}+{\bf G}^{\rm qs}$),
grey dashed (using ${\bf G}^{\rm B}+{\bf G}^{\rm f}$), and black solid (using ${\bf G}^{\rm far}+{\bf G}^{\rm back}_{1}$); circles indicate  full-dipole results at three selected locations (around 10~nm, 6~nm, and 2~nm), showing excellent agreement
with results obtained using ${\bf G}^{\rm B}+{\bf G}^{\rm f}+{\bf G}^{\rm qs}$. Note that the curves are somewhat jagged as we have only evaluated the Green functions at select locations.}
\label{fig:A}
\end{figure}

{To elaborate on the approximation made to Eq.~(\ref{eq:2_App}), we note that $\mG^\text{others}$ includes also a first-order background term, }
\begin{align}
{\bf G}_1^{\rm back}({\bf r}_1,{\bf r}_2;\omega)
= \int_{V}{\bf G}^{\rm B}({\bf r}_1,{\bf r'};\omega)\cdot\Delta\varepsilon({\bf r'},\omega){\bf G}^{\rm B}({\bf r'},{\bf r}_2;\omega)d{\bf r'},
\label{eq:6_App}
\end{align}
which is negligible above about 40~nm, and finite below this value (black solid curve). {This term, however, is only a first-order Born approximation and turns out  to be not reliable. 
To make this clear}, in Fig.~\ref{fig:A}(b), we compare results with this term against the proposed ${\bf G}^{\rm qs}$ (i.e., the nonperturbative half space solution), and only the latter is seen to be in agreement with full dipole FDTD results \cite{FDTD}; indeed, the agreement is quantitative. Although we have chosen a single frequency here, in the main text we also show excellent agreement as a function of frequency at the spatial location of only 2~nm.

{It is interesting to note that} Sauvan \emph{et al.}~\cite{Sauvan13} also found that a single QNM expansion works well for {evaluating the SE rate at a range of spatial distances outside the MNP
using a dipole emitter position-dependent effective mode volume in the Purcell formula, and adding to it a factor of unity, seemingly ``by hand.'' The formalism presented here: $(i)$ defines a mode volume that is characteristic of the surface plasmon mode alone, and a separate factor $\eta$ that characterizes the dipole coupling to that mode, $(ii)$ provides an explanation for the factor of unity, $(iii)$ provides an efficient and intuitive method for accurately calculating both SE rates and field propagation effects to arbitrarily large distances from the MNP, and $(iv)$ also works accurately at short distances where Ohmic loss mechanisms are important.

\section{Peak spontaneous emission
deviation factor}
\label{AppC}
In the main text, we defined the relative SE rate in terms of the Purcell factor 
\begin{align}
F_a({\bf r}_a,\omega) = F_\text{P}\,\eta(\mr_a,{\mn}_a;\omega) +1,
\label{Eq:GeneralizedPurcell_App}
\end{align}
where we stated that
 $\eta(\mr_a,{\mn}_a;\omega)$ 
accounts for any deviations at ${\bf r}_a$ from the field maximum $\mr_0$, cavity polarization, and cavity resonant frequency.  To see how this comes about, we write the SE rate, for a 3D geometry, as
\begin{align}
F_a(\mr_a,\omega) &= \frac{6\pi c^3}{n_\text{B}\omega^3}\text{Im}\left\{\mathbf{n}_a\cdot
\mG^\text{far}(\mr_a,\mr_a;\omega)\cdot\mathbf{n}_a\right\}, \nonumber \\
& = 1 +  \frac{6\pi c^3}{n_\text{B}\omega}
\text{Im}\left\{ \mathbf{n}_a\cdot 
\frac{ \mF_\mu(\mr_a,\omega)\mF_\mu (\mr_a,\omega) }{ 2\tlo_\mu(\tlo_\mu-\omega )}
\cdot\mathbf{n}_a\right \},   
  \nonumber \\
& = 1 + \frac{3Q}{4\pi^2}\left(\frac{\lambda_\text{B}}{n_\text{B}}\right)^3\frac{\omega_\text{c}^2\gamma_\text{c}}{\omega}
\text{Im}\left\{\mathbf{n}_a\cdot
\frac{\varepsilon_\text{B}\mF_\mu(\mr_a,\omega)\mF_\mu(\mr_a,\omega)}{\tlo_\mu(\tlo_\mu-\omega)}
\cdot\mathbf{n}_a\right\}. \nonumber \\
\label{eq:final}
\end{align}
Combining this with Eq.~(\ref{Eq:GeneralizedPurcell_App}), 
 we derive
the peak SE deviation factor,
\begin{align}
\eta(\mr_a,\mathbf{n}_a;\omega) = 
\frac{V_{\rm eff}\omega_c^2\gamma_c}{\omega}\,
\text{Im}\left\{\mathbf{n}_a\cdot
\frac{\varepsilon_\text{B}\mF_\mu(\mr_a,\omega)\mF_\mu(\mr_a,\omega)}{\tlo_\mu(\tlo_\mu-\omega)}\cdot
\mathbf{n}_a\right\}.
\end{align}
For spatial points near the resonator,  then one can simply replace $\tilde{\bf F}_{\mu}({\bf r}_a)$ with $\tilde{\bf f}_{\mu}({\bf r}_a)$.
Note, to derive these expressions we have used the Green function of a 3D homogeneous medium; expressions for ${\bf G}^{\rm B}$ for 2D and 3D are given in, e.g., Ref.~\cite{G2D} and Ref.~\cite{NovotnyAndHecht_2006}, respectively, so the 2D expressions can be derived in  the same way.\\

\section{Effective mode volumes: correspondence with recently published results}
\label{AppA}
In a recent paper,~\cite{Sauvan13} Sauvan \emph{et al.} address the problem of normalization of QNMs in  dispersive resonators. In particular, the authors use the Lorentz reciprocity theorem to derive a generalized mode volume for use in the Purcell formula. 
Below, we show that the normalization in Ref.~\cite{Sauvan13} is a generalization of a previous result by Leung \emph{et al.}~\cite{Leung941} to the case of vector fields and magnetic materials. For non-magnetic, dispersionless materials, the corresponding mode volume reduces to  the same generalized mode volume that was {previously} introduced for leaky optical cavities~\cite{Philip:OL12}. 
Sauvan~\emph{et al.} define a generalized mode volume as\cite{Sauvan13}
\begin{align}
V = \frac{\int_V \left [\mE_\text{c} \cdot\frac{\partial(\omega\varepsilon_0\epsilon_\text{})}{\partial\omega}\mE_\text{c}  - \mH_\text{c} \cdot\frac{\partial(\omega\mu_\text{})}{\partial\omega}\mH_\text{c} \right ]\ud\mr}{2\varepsilon_0\varepsilon_\text{}(\mr_0)\{\mE_\text{c}(\mr_0)\cdot\mathbf{u}\}^2},
\label{Eq:SauvanModeVol}
\end{align}
where $\mr_0$ and $\mathbf{u}$ denote the position of the emitter and the direction of the emitter dipole moment, respectively, and the integration volume is over all space. Here $\varepsilon_0$ and $\mu_0$ denote the permittivity and permeability of free space, respectively, and $\varepsilon_\text{}=\varepsilon_\text{}(\mr,\omega)$ and $\mu_\text{}=\mu_\text{}(\mr,\omega)$ are the relative permittivity and permeability describing the resonator. 
The fields $\mE_\text{c}=\mE_\text{c}(\mr,\tlo_\text{c})$ and $\mH_\text{c}=\mH_\text{c}(\mr,\tlo_\text{c})$ are the resonant electric and magnetic field cavity modes {with complex eigenfrequency $\tilde\omega_c$. These modes are solutions to the wave equation with the outgoing wave boundary condition and are related as}
\begin{align}
\nabla\times\mE_\text{c} &= i\omega_\text{c}\mu_0\mu_\text{}\mH_\text{c} , \label{Eq:curlE}\\
\nabla\times\mH_\text{c} &= -i\omega_\text{c}\varepsilon_0\varepsilon_\text{}\mE_\text{c}\label{Eq:curlH}.
\end{align}
{In this article, we work only with the QNMs of the electric field that we denote $\mft_\mu(\mr)$.} For scalar fields, the outgoing wave boundary condition is known as the Sommerfeld radiation condition. The generalization to electromagnetic vector fields is the Silver-M\"uller radiation condition which may be written as~\cite{Martin_MultipleScattering}
\begin{align}
\frac{\mathbf{r}}{r}
\times\nabla\times\mE + \text{i}k\mE \rightarrow 0\quad\text{as}\,\,r\rightarrow\infty,
\label{Eq:SilverMullerCond}
\end{align}
where $k=n_\text{B}\omega/c$ is the magnitude of the wave vector in the homogeneous material with refractive index $n_\text{B}$, and  $r=|\mr|\rightarrow\infty$ 

To see the correspondence with previously published results on non-magnetic materials, we set $\mu_\text{}=1$ and focus on the numerator in Eq.~(\ref{Eq:SauvanModeVol}) (in which we include the factor of $2\varepsilon_0$ from the denominator) and show how it relates to the inner product of Ref.~\cite{Leung941}. Using Eq.~(\ref{Eq:curlE}), we {first} rewrite the numerator as
\begin{align}
\langle\langle\mE_\text{c}|\mE_\text{c}\rangle\rangle_\text{Sauvan} &= \frac{1}{2\varepsilon_0}\int_V \left [\varepsilon_0\frac{\partial(\omega\varepsilon_\text{})}{\partial\omega} \mE_\text{c}\cdot\mE_\text{c}  - \mu_0\left(\frac{-\text{i}}{\mu_0\tlo_\text{c}}\right)^2\left(\nabla\times\mE_\text{c}\right)\cdot\left(\nabla\times\mE_\text{c}\right)
\right ]\ud\mr.
\label{Eq:innerProdSauvan}
\end{align}
Now we use the vector generalization of Green's 
identity of the first kind~\cite{Martin_MultipleScattering},
\begin{align}
\int_V \left [ (\nabla\times\mathbf{P})\cdot(\nabla\times\mathbf{Q}) - \mathbf{P}\cdot\nabla\times\nabla\times\mathbf{Q}
\right ]\,\ud\mr = \int_S\mathbf{n}\cdot(\mathbf{P}\times\nabla\times\mathbf{Q})\,\ud\mathbf{a},
\end{align}
to rewrite the integral as
\begin{align}
\langle\langle\mE_\text{c}|\mE_\text{c}\rangle\rangle_\text{Sauvan} = &\frac{1}{2}\int_V
\left [
\frac{\partial
(\omega\varepsilon_\text{})}{\partial\omega} \mE_\text{c}\cdot\mE_\text{c}
+\frac{c^2}{\tlo^2_\text{c}}\mE_\text{c}\cdot\nabla\times\nabla\times\mE_\text{c}\right ]
\ud\mr \nonumber \\
& +\frac{1}{2}\frac{c^2}{\tlo^2_\text{c}} \int_S \mathbf{n}\cdot\left( \mE_\text{c}\times \nabla\times\mE_\text{c} \right)\ud\mr.
\end{align}
Last, using Eqs.~(\ref{Eq:curlE})-(\ref{Eq:SilverMullerCond}) and the identity $\mathbf{A}\cdot\left(\mathbf{B}\times\mathbf{C}\right) 
= -\mathbf{B}\cdot\left(\mathbf{A}\times\mathbf{C}\right)$, we find that
\begin{align}
\langle\langle\mE_\text{c}|\mE_\text{c}\rangle\rangle_\text{Sauvan} &= \frac{1}{2}\int_V \left( \frac{\partial(\omega\varepsilon_\text{})}{\partial\omega} + \varepsilon_\text{}\right) \mE_\text{c}\cdot\mE_\text{c}\ud\mr +\text{i}\frac{\sqrt{\varepsilon_\text{}}\,c}{2\tlo_\text{c}} \int_S \mE_\text{c}\cdot\mE_\text{c}\ud\mr. \\
&= \int_V \sigma(\mr,\tlo)\mE_\text{c}\cdot\mE_\text{c}\ud\mr +\text{i}\frac{n_\text{B}c}{2\tlo_\text{c}} \int_S \mE_\text{c}\cdot\mE_\text{c}\ud\mr,
\label{Eq:LeeInnerProdWithDisp}
\end{align}
where 
\begin{align}
\sigma(\mr,\omega)=\frac{1}{2\omega}\frac{\partial(\varepsilon_\text{}\omega^2)}{\partial\omega}.
\end{align}
This shows directly that Eq.~(\ref{Eq:innerProdSauvan}) 
is the generalization to vector fields in general dispersive and magnetic material systems of the inner product in Ref.~\cite{Leung941}, and for non-magnetic and dispersionless materials Eq.~(\ref{Eq:SauvanModeVol}) reduces to the {exact same generalized mode volume} that was previously introduced 
{in Ref.~\cite{Philip:OL12}}. 

Sauvan \emph{et al.} argue that the use of Perfectly Matches Layers (PMLs) are an important and intricate part of the formulation of the mode volume, since it leads to a finite value for the inner product in Eq.~(\ref{Eq:innerProdSauvan}) despite the divergence of the QNMs at large distances from the cavity. The equivalence between the norms, as discussed above, would then suggest that use of PMLs are necessary also for the evaluation of Eq.~(\ref{Eq:LeeInnerProdWithDisp}). As explained in the supplementary information to Ref.~\cite{Sauvan13}, however, the integrand in Eq.~(\ref{Eq:SauvanModeVol}) is an invariant of the coordinate transforms that can be calculated with any choice of PMLs. This means in particular that the integrand is invariant under the trivial operation where no coordinate transform is performed, and therefore the integral must be convergent to begin with. In other words, the integrand in Eq.~(\ref{Eq:SauvanModeVol}) cannot be divergent and therefore must tend to zero at large distances even though each of the fields $\mE_\text{c}$ and $\mH_\text{c}$ do in fact diverge as $r\rightarrow\infty$. In a similar way, the sum of the two terms in Eq.~(\ref{Eq:LeeInnerProdWithDisp}) remain finite as a function of integration domain size, even though each of the terms diverge. {This convergence is shown explicitly in Figs.~\ref{fig:2}(b) and \ref{fig:4}(b). }

\section*{References}


\end{document}